\documentstyle[floats,aps,prl,epsf,twocolumn]{revtex}

\begin{document}
\draft

\wideabs{

\title{$^{63}$Cu NMR Evidence for Enhanced Antiferromagnetic
Correlations around Zn Impurities in YBa$_2$Cu$_3$O$_{6.7}$}

\author{M.-H. Julien$^{1,2}$, T. Feh\'er$^{1,3}$,
M. Horvati\'c$^{1}$, C. Berthier$^{1,2}$, O.N.~Bakharev$^{1,4}$,
P. S\'egransan$^{2}$, G.~Collin$^{5}$~and~J.-F.~Marucco$^{6}$}

\address{$^1$Grenoble High Magnetic Field Laboratory, CNRS and MPI-FKF,
BP 166, F-38042 Grenoble Cedex 9, France}
\address{$^2$Laboratoire de Spectrom\'etrie Physique, Universit\'e
J. Fourier, BP 87, F-38402 St. Martin d'H\`eres, France}
\address{$^3$Physics Institute, Technical University
of Budapest, H-1521, POB. 91, Budapest, Hungary}
\address{$^4$Magnetic Resonance Laboratory, Kazan State University, 420008 Kazan, Russia}
\address{$^5$Laboratoire L\'eon Brillouin, Centre d'Etudes de Saclay,
CEA-CNRS, 91191 Gif sur Yvette, France}
\address{$^6$Laboratoire des Compos\'es non Stoechiom\'etriques,
Universit\'e Paris-Sud, 91405 Orsay, France}

\date{\today}

\maketitle


\begin{abstract}
Doping the high-$T_c$ superconductor YBa$_2$Cu$_3$O$_{6.7}$ with
1.5 \% of {\it non-magnetic} Zn impurities in CuO$_2$ planes is
shown to produce a considerable broadening of $^{63}$Cu NMR
spectra, as well as an increase of low-energy magnetic
fluctuations detected in $^{63}$Cu spin-lattice relaxation
measurements. A model-independent analysis demonstrates that these
effects are due to the development of staggered magnetic moments
on many Cu sites around each Zn and that the Zn-induced moment in
the bulk susceptibility might be explained by this staggered
magnetization. Several implications of these {\it enhanced}
antiferromagnetic correlations are discussed.

\end{abstract}

\pacs{PACS numbers: 76.60.-k, 74.25.Ha, 74.62.Dh}

}

\narrowtext


The precise nature of antiferromagnetic (AF) correlations and how
they influence electronic properties are the most puzzling aspects
of high-$T_c$ cuprate superconductors. A piece of this puzzle is
the substitution of Cu$^{2+}$ ions ($S$=1/2) by dilute impurities,
such as Zn, which is known to suppress spectacularly $T_c$ (about
20 K/\% of Zn in the underdoped regime), and to localize charges
\cite{transport}. The most dramatic effects of Zn doping concern
the magnetism of CuO$_2$ planes \cite{Alloul99}: although Zn is
non-magnetic, the observation of specific EPR and NMR resonances
have shown that the Curie term in the bulk susceptibility of
Zn-doped samples \cite{Mendels99} is due to a magnetic moment
induced on Cu sites around Zn
\cite{Finkelstein90,Mahajan94,Williams95,Ishida96}. Furthermore,
the broadening of $^{63}$Cu~\cite{Walstedt93,Zheng} and $^{17}$O
NMR lines \cite{Bobroff97} reveals a distribution of moment
magnitudes, which has been attributed to a spatially inhomogeneous
spin polarization extending over several lattice sites around Zn.
Both the symmetric character and the temperature ($T$) dependence
of the line broadening can be accounted for by staggered spin
density oscillations, as expected from the AF character of
magnetic correlations \cite{Walstedt93,Bobroff97,Morr98}.
Zn-doping also has drastic effects on spin dynamics: low-energy AF
fluctuations probed by neutron scattering are enhanced
\cite{Neutrons,Bourges96}, and spin-freezing eventually occurs
\cite{Neutrons,Bourges96,Mendels94,Bernhard98}. From a general
point of view, such clear manifestations of Cu$^{2+}$ moments
recall that metallic cuprates are primarily doped
antiferromagnets. Nevertheless, a clear understanding of these
effects, particularly the origin of the magnetic moment, remains
elusive.

Here, we report on a $^{63}$Cu~NMR study of
YBa$_2$(Cu$_{0.99}$Zn$_{0.01}$)$_3$O$_{6.7}$. By exploiting the
peculiarities of hyperfine interactions for Cu nuclei in CuO$_2$
planes, these measurements provide a model-independent
demonstration that the local magnetization becomes staggered
around Zn impurities. The uncompensated sum of the staggered
moments is found to be of the same order of magnitude as the
moment detected in bulk measurements. Additional low-energy spin
fluctuations from these staggered regions are identified within
the pseudogap through measurements of the nuclear spin-lattice
relaxation rate (1/$T_1$). The physical picture implied by these
results is that AF correlations are {\it enhanced}, not destroyed,
around impurities.

The measurements were~performed~on~a single crystal of
YBa$_2$(Cu$_{0.99}$Zn$_{0.01}$)$_3$O$_{6.7}$ of $\sim$400~mg
weight. SQUID measurements have shown that $T_{c}$=38 K, but it is
reduced to $\sim$20 K or less in magnetic fields ($H$) of 12-15~T
used here. The $T_{c}$ value is in agreement with published data
\cite{Mendels99} for 1\% Zn/Cu ({\it i.e. 1.5\% Zn/Cu(2)}).

\begin{figure*}[!t]
\centerline{\epsfxsize=170mm \epsfbox{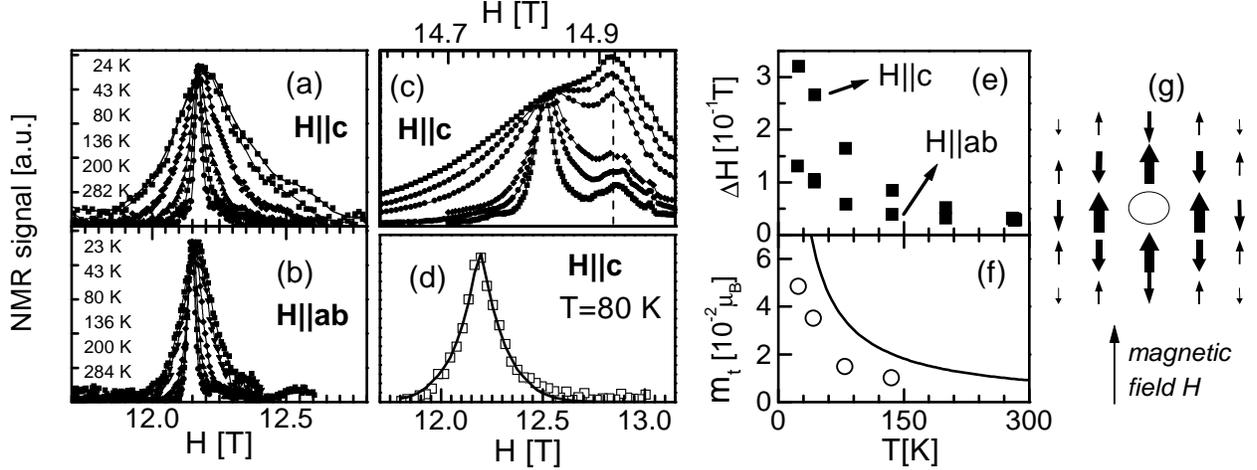}}
\vspace{-4.8cm} \caption{ (a) Normalized field swept $^{63}$Cu(2)
spectra with $H$$||$$c$, the Cu(1) signal being suppressed by a
differential spin-echo sequence. (b) $^{63}$Cu(2) spectra obtained
by the same method, but $H$$||$$ab$. (c) $^{63}$Cu spectra for
$H$$||$$c$ without the differential sequence ($T$=26, 49, 83, 156,
203 and 285 K); the line shows the $^{63}$Cu(1) signal;
intensities are normalized at the Cu(2) peak position. (d) Fit of
a $^{63}$Cu(2) spectrum using the ${\cal I}$ function explained in
the text. (e) Width at half maximum of the $^{63}$Cu(2) spectra
from plots (a,b). (f) Circles: total magnetization per Zn carried
by staggered moments (see text). Line: magnetization for an
effective moment $p_{\rm eff}$=1$\mu_B$ at 12.2 T. (g) Schematic
drawing of staggered moments (arrows), induced on a characteristic
length $\sim$$\xi$ around a single Zn impurity (circle). The size
of each arrow represents the value of the moment. }
\end{figure*}

Fig. 1(c) shows typical $^{63}$Cu~NMR spectra for $H$$\Vert
$$c$-axis. Each spectrum consists of two broad lines, associated
with the plane site Cu(2) and the chain sites Cu(1) (the various
inequivalent Cu(1) sites are not resolved here). Spectra in Fig.
1(a,b) are obtained with a differential spin-echo pulse sequence
designed to suppress the Cu(1) signal, which has longer
spin-lattice ($T_{1}$) and spin-spin ($T_{2}$) relaxation times
than Cu(2). The Cu(1) signal is not totally suppressed, but the
mixing with the Cu(2) signal is largely reduced, allowing direct
fits of the Cu(2) lineshape. In Fig.1, the amplitudes at the
center of the Cu(2) line are normalized to a common value for all
temperatures. This makes very apparent the huge broadening of the
Cu(2) line as $T$ is reduced. On the other hand, the width of the
Cu(1) line is almost $T$-independent. This shows that Zn
substitutes only the planar Cu site, not the chain site.

Zn-induced magnetic broadening NMR spectra (already reported for
$^{63}$Cu in \cite{Walstedt93,Zheng}) indicates a distribution of
hyperfine fields $h_i$. The NMR spectrum simply corresponds to the
histogram of this distribution. For $H$$\|$$\alpha$=$c$,$ab$, in
the paramagnetic phase, $h_i$ at a site ($r_i$) comes from an
orbital term, which is $T$-independent, plus a spin term due to an
anisotropic coupling ($A_\alpha^{\rm onsite}$) to the on-site
Cu$^{2+}$ moment and to an isotropic transferred coupling ($B$) to
the four Cu$ ^{2+}$ first neighbors ($\epsilon$) \cite{hyperfine}:
\begin{equation}
h_{i}=h_{{\rm orb}}+A_{\alpha }^{\rm onsite}\langle S_{z}^{i}\rangle +B\sum_{\epsilon
=1..4}\langle S_{z}^{i+\epsilon }\rangle
\end{equation}
The spin component can be expressed in $q$-space as:
\begin{equation}
h_{i}^{\rm spin}=\frac{1}{N}\sum_{\overrightarrow{q}}A(\overrightarrow{q}%
)\left\langle S_{z}(\overrightarrow{q})\right\rangle
\exp \left(i\overrightarrow{q}.\overrightarrow{r_{i}}\right)
\end{equation}
where $A_\alpha(\overrightarrow{q})=A_{c(ab)}^{\rm
onsite}+2B\cos(q_{x}a) +2B\cos(q_{y}b)$. A well-established NMR
result in YBa$_{2}$Cu$_{3}$O$_{6+x}$ (YBCO) is that
$A_{c}(q=0)=A_{c}^{\rm onsite}+4B\simeq 0$, which means that
$h_{i}^{\rm spin}\simeq 0$ when the time averaged spin
polarization is uniform in space, while $A_{c} $ is maximum for
$\overrightarrow{q}=\overrightarrow{Q}_{AF}=(\pi /a$, $\pi /a)$.
The huge, $T$-dependent, distribution of $h_{i}$ observed here is
thus a clear indication that the local magnetization is spatially
strongly modulated $\langle S_{z}^{i}\rangle           \,
{\approx} \!\!\!\! / \; \langle S_{z}^{i+\epsilon}\rangle$. Since
it has to be so on the length of one lattice spacing, any
non-microscopic modulation, such as induced by doping
inhomogeneities, is ruled out. In addition, since $A_c^{\rm
onsite}$ and $B$ have opposite signs, large values of $h_i$ are
obtained when $\langle S_{z}^{i}\rangle$ and $\langle
S_{z}^{i+\epsilon}\rangle$ have opposite signs, {\it i.e.} if the
magnetization is~{\it staggered}.

Further confirmation comes from measurements with
$H$$\|$$ab$-plane. In this orientation, the $^{63}$Cu(2) NMR
spectrum also broadens on cooling, but, unexpectedly, its width
becomes less than for $H$$\|$$c$ by a factor 2.6$\pm$0.2 at
low-$T$ (Fig. 1e). Since the ratio $A_{c}(q)/A_{ab}(q)$ is close
to 2.6 in the vicinity of $\overrightarrow{Q}_{AF}$, this confirms
without any detailed model that the staggered component of the
magnetization is dominant.

This conclusion can be checked quantitatively by computer
simulations of the lineshape according to the following model (see
also \cite{Walstedt93,Bobroff97,Morr98}). Any localized magnetic
perturbation at a site $r_{\rm imp}$ is expected to polarize
surrounding electrons $r_i$ with a magnitude determined by the
real-space spin susceptibility $\chi^\prime(r_i-r_{\rm imp})$
\cite{White}. This latter is the Fourier transform (FT) of the
real part of $\chi(q)$, which in underdoped YBCO is well-known to
be peaked at, or near, $Q_{AF}$ as a consequence of AF
correlations. Around a single impurity, the spin polarization at a
site $r_{i}$ is given by $h^{\rm spin}_i=\frac{1}{N}\lambda
\sum_{q}\chi^\prime(q)A(q) \exp
[i\overrightarrow{q}.(\overrightarrow{r_{i}}-\overrightarrow{r}_{\rm
imp})]$, in which we constrain $\sum_{q}\chi^\prime (q)=0$ to
impose the absence of moment at the Zn site, and
$\lambda$=$\chi^\prime(Q_{AF}) V_{\rm eff}$, where $V_{\rm eff}$
stands for the local effective perturbating potential. Assuming
that the response to the randomly distributed impurities are
additive, the local field distribution in the plane $h^{\rm
spin}_i$ is just given by the FT of $\lambda \chi^\prime (q){\cal
D}(q)A(q)$, where ${\bf {\cal D}(q)}$ is the inverse FT of a
random distribution of impurity sites ${\bf
\sum_{\overrightarrow{r}}}_{\rm imp}\delta
(\overrightarrow{r}-\overrightarrow{r}_{\rm imp})$, taken on a
256$\times$256 lattice. The histogram of the $h^{\rm spin}_i$
convoluted with an intrinsic linewidth, which is small compared to
the broadening, gives the NMR lineshape.

Three different models for $\chi^\prime(q)$ have been considered:
a commensurate Lorentzian ${\cal L}(\xi,q)= 1/(1+\xi
^{2}(\overrightarrow{q}-\overrightarrow{Q}_{AF})^{2})$, a
commensurate Gaussian ${\cal G}(\xi,q)$=$\exp [-\xi
^{2}(\overrightarrow{q}-\overrightarrow{Q}_{AF})^{2}/2$], and a
four peaks incommensurate function ${\cal I}(\xi ,\delta ,q)$
corresponding to the functional form used to model recent neutron
data in YBa$_{2}$Cu$_{3}$O$_{6.6}$ \cite{Mook98}. Spectra for
$H$$||$$c$ and 25~K$\leq$$T$$\leq$135~K were first fitted using a
non linear procedure letting free $\lambda$ and $\xi$. For both
${\cal L}$ and ${\cal G}$ functions, the best fits are obtained
for $\xi/a\simeq$2.5 - 4. Using the values of $\lambda$ and $\xi$
obtained from these fits, and $g_c/g_{ab}$=1.15 \cite{Walstedt92},
one reproduces the observed anisotropy ($\sim$2.6) for the low-$T$
linewidth \cite{nn}. The function ${\cal I}(\xi,\delta ,q)$ gives
similar results. If $\delta$ is also set as a fitting parameter,
the best fit is obtained for $\delta $=0.26$\pm$0.01 r.l.u. and
$\xi /a$=5$\pm$1, independent of $T$ within error bars in the
range 25~K$\leq$$T$$\leq$80~K. Note that a roughly $T$-independent
$\xi$ is not unexpected below $\sim$120~K \cite{Takigawa94}, so
our results do not conflict with the variation of $\xi$ inferred
at higher $T$ \cite{Bobroff97,Morr98}.

The remarkable agreement between the above values and neutron data
\cite{Neutrons,Mook98} demonstrates that the essential features of
the response to Zn doping are included in the model (note that
slight differences in lineshapes are obtained with ${\cal L}$,
${\cal G}$ and ${\cal I}$ functions, but they are not sufficient
to favor one model). It also gives us confidence that the obtained
$\lambda$ values are reasonable. This allows for the first time to
estimate the total magnetization carried by the staggered
response: $m_{\rm tot}=\sum_{i}\langle S_{z}^i\rangle$. Although
one might naively expect $m_{\rm tot}\sim 0$ for a staggered
magnetization, not only $m_{\rm tot}$ is nonzero, but it is
comparable to the bulk magnetization $M_{\rm bulk}=p^2_{\rm
eff}H/3k_B T$ with $p_{\rm eff}=1\mu_B$ \cite{Mendels99}, in the
case of ${\cal L}$ and ${\cal I}$ functions (Fig 1f). The ${\cal
G}$ function, which is less likely
\cite{Alloul99,Bobroff97,Morr98}, gives about three times smaller
$m_{\rm tot}$, because the amplitude on the nearest neighbors
($nn$) of the impurity is smaller than that obtained with ${\cal
L}(q)$ models. The accuracy of the comparison between $m_{\rm
tot}$ and  $M_{\rm bulk}$ is further limited by several
uncertainties: value of $M_{\rm bulk}$, remaining Cu(1) signal,
intensities possibly biased by inhomogeneous relaxation times, as
well as a possible deviation from the asymptotic model on the four
$nn$.

Despite these uncertainties, the results obtained with Lorentzian
models of $\chi(q)$ suggest that the bulk magnetization ($i.e.$
the "Zn-induced moment") may be explained as the sum of staggered
moments only. It is important to realize that this is conceptually
different from an approach of Zn doping where the staggered
response is attributed to the polarization of AF-correlated "band"
electrons by a putative moment on the four Cu $nn$ of a Zn. In our
view, the staggered magnetization, which must occur in response to
any kind of defect \cite{rem}, naturally leads to such a large
moment on $nn$. In other words, Zn only reveals already-existing
moments, localized on Cu sites of the doped antiferromagnet (at
least in the underdoped regime).

As a matter of fact, it must be emphasized that the enhanced
staggered magnetization makes clear that, counter-intuitively, AF
correlations are not destroyed but {\it enhanced} by Zn.

It is interesting to make a connection with recent studies of
non-magnetic impurities in AF quantum spin chains. There, the
staggered magnetization inferred from NMR is very well-explained
by the same kind of model as used here, except for the appropriate
$\chi(q)$ \cite{exp1D}. Such enhancement of AF correlations has
attracted considerable theoretical interest because it was also
shown to be a property of the "spin-liquid" ground state
\cite{theo1D}. Enhanced moments around vacancies are also found in
some approaches of the Heisenberg or $t$-$J$ models in 2D
\cite{theo2D}. However, the Zn-induced staggered magnetization,
seen at relatively high $T$ in a paramagnetic state, is expected
from the general arguments used above. So, it is not by itself an
indication of spin-liquid ground state.


It is natural to expect that staggered moments affect dynamical
properties, at least locally. We have measured the
$^{63}$Cu~spin-lattice relaxation rate (1/$T_1$), which is a probe
of the dynamical susceptibility at low energy:
%
\begin{equation}
\frac{1}{T_1} = k_{\rm B} T\;\;^{63}\gamma^2
\sum_{q,\alpha\perp H} \frac{|A_\alpha(q)|^2}{(g_{\alpha}\mu_{\rm B})^2}
\;\;\frac{\chi^{\prime\prime}(q,\omega_n)}{\omega_n}
\end{equation}
The recovery law of the nuclear magnetization $M_z(t)$ following
an inversion pulse was fitted with two contributions at all
temperatures: a short component due to $^{63}$Cu(2) nuclei and a
longer one from $^{63}$Cu(1).

The $T$-dependence of $(T_1T)^{-1}$ is shown in Fig. 2, along with
data from Ref. \cite{Takigawa91} in Zn-free
YBa$_{2}$Cu$_3$O$_{6.63}$. Although similar in the range 300-150
K, the data show opposite trends below $\sim$100~K: $(T_1T)^{-1}$
decreases with $T$ in Zn-free YBCO, this is the well-known
pseudogap behaviour, while $(T_1T)^{-1}$ increases in YBCO with
Zn, signaling an enhancement of $\chi"(Q,\omega_n)$. Still, the
plateau of $(T_1T)^{-1}$ around 120~K for YBCO with Zn is
strikingly reminiscent of the maximum at the characteristic
temperature $T^*$$\simeq$140~K, which is commonly used to define
the pseudogap energy scale. Although some decrease of $T^*$, as
found in \cite{Zheng}, is not excluded by our data, the pseudogap
(for AF excitations) appears to be filled-in by Zn-induced
excitations, rather than destroyed. This is in perfect agreement
with the neutron scattering data at $q$=$Q$ plotted in the inset
to Fig. 2 \cite{Bourges96} (see also \cite{Neutrons}). Most
likely, the pseudogap remains intact far from Zn where the
magnetization is small. Note that this finding is complementary
to, but different from the fact that the decrease of
$\chi^\prime$($q$=0,$\omega$=0) between 300 and 100 K is
unaffected by Zn at sites far from Zn \cite{Mahajan94}. Again
there is an analogy with recent work in 1D: for example, the
spin-gap of the Zn-doped ladder SrCu$_2$O$_3$ coexists with
low-energy AF fluctuations \cite{Azuma98}. We also mention that
our $T_1$ data are rather qualitative as they do not allow a
quantitative analysis of Zn-induced $^{63}T_1$ contributions.
However, a recent extraction of these contributions for Ni-doping
led to conclusions similar to ours \cite{Itoh99}.

\begin{figure}[!t]
\vspace{-9mm}
 \epsfxsize=88mm
 $$\epsffile{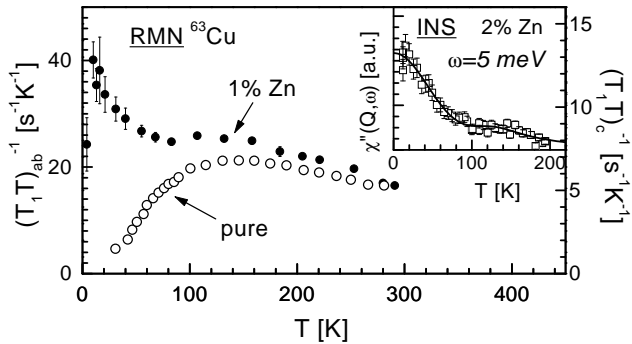}$$
\vspace{-8.5cm} \caption{$^{63}(T_1T)^{-1}$ data in in
YBa$_2$(Cu$_{0.99}$Zn$_{0.01}$)$_3$O$_{6.7}$ ($H$$||$$ab$, left
scale) and in YBa$_{2}$Cu$_3$O$_{6.63}$ ($H$$||$$c$, right scale,
from \protect\cite{Takigawa91}). Inset: neutron scattering data in
YBa$_2$(Cu$_{0.99}$Zn$_{0.02}$)$_3$O$_{6.7}$ from Ref.
\protect\cite{Bourges96}. }
\end{figure}


A remaining puzzle is why the bulk susceptibility obeys a pure
Curie law ($\chi\sim (T+\theta)^{-1}$, $\theta\simeq$0)
\cite{Mendels99}, as if the moments were not interacting, even at
concentrations where the mean distance between impurities is much
less than 2$\xi$ the typical diameter of staggered areas. Most
probably, individual spins in a staggered area are locked to each
other, so each area behaves as a single moment, but negligible
interaction between them is surprising. Actually, a pure Curie
susceptibility is also found in the "cluster spin-glass" phase of
La$_{2-x}$Sr$_x$CuO$_4$ \cite{Chou95}, where staggered domains are
thought to be encircled by doped-holes \cite{Julien99}. We
speculate that analogous hole segregation in regions of small
staggered magnetization ({\it i.e.} far from Zn) may largely
reduce interactions between "AF clusters". This is consistent with
the suppression of the superfluid density in areas of
characteristic length 2$\xi$ \cite{Nakano98,Nachumi96}.
Eventually, the tendency of doped holes to avoid AF regions around
Zn, where their mobility would be reduced, could rationalize
anomalous charge localization effects \cite{transport}, including
stripe pinning. Subsequent spin-freezing is expected in most
cases.

Finally, we stress that the Zn-induced staggered magnetization
does {\it not} conflict with the possible presence of moments in
impurity-free cuprates \cite{Bernhard98}: there are evidently
staggered moments at low hole doping \cite{Julien99}. Moreover,
the decrease of $T_c$ when AF correlations are enhanced does not
mean that antiferromagnetism and superconductivity are unrelated.
Rather, the Zn-induced static character of AF correlations and/or
randomness are presumably detrimental to high-$T_c$.

We thank R. Calemczuk and C. Marin for $T_c$ measurements and H.
Alloul, J. Bobroff, P. Mendels, A. J\'anossy, G. Aeppli and F.
Tedoldi for discussions. T.F. was supported by a Hungarian State
Grant OTKA T029150.



\begin{references}
\bibitem{nn} Although we cannot ascertain that the four Cu $nn$
of a Zn are observed, this does not affect significantly the fit
results, since these nuclei contribute a small intensity in the
tails of the line.
\end{references}
\end{document}